\documentclass[apj]{emulateapj}

\usepackage{url}
\usepackage{wasysym}

\begin{document}

\title{The Interaction of Venus-like, M-dwarf Planets with the Stellar Wind of Their Host Star}

\author{O. Cohen\altaffilmark{1}, Y. Ma\altaffilmark{2}, J.J. Drake\altaffilmark{1}, A. Glocer\altaffilmark{3}, C. Garraffo\altaffilmark{1},
J.M. Bell\altaffilmark{4}, and T.I. Gombosi\altaffilmark{5}}

\altaffiltext{1}{Harvard-Smithsonian Center for Astrophysics, 60 Garden St. Cambridge, MA 02138, USA}
\altaffiltext{2}{Institute of Geophysics and Planetary Physics, University of California, Los Angeles, Los Angeles, California USA}
\altaffiltext{3}{NASA/GSFC, Code 673 Greenbelt, MD 20771, USA.}
\altaffiltext{4}{Center for Planetary Atmospheres and Flight Sciences, National Institute of Aerospace, Hampton, VA 23666, USA}
\altaffiltext{5}{Center for Space Environment Modeling, University of Michigan, 2455 Hayward St., Ann Arbor, MI 48109, USA}

\begin{abstract} 

We study the interaction between the atmospheres of Venus-like, non-magnetized exoplanets orbiting an M-dwarf star, and the stellar wind using a multi-species Magnetohydrodynaic (MHD) model. We focus our investigation on the effect of enhanced stellar wind and enhanced EUV flux as the planetary distance from the star decreases. Our simulations reveal different topologies of the planetary space environment for sub- and super-Alfv\'enic stellar wind conditions, which could lead to dynamic energy deposition in to the atmosphere during the transition along the planetary orbit. We find that the stellar wind penetration for non-magnetized planets is very deep, up to a few hundreds of kilometers. We estimate a lower limit for the atmospheric mass-loss rate and find that it is insignificant over the lifetime of the planet. However, we predict that when accounting for atmospheric ion acceleration, a significant amount of the planetary atmosphere could be eroded over the course of a billion years.
  
\end{abstract}



\section{INTRODUCTION}
\label{sec:Intro}

Nearly two decades after the first detections of exoplanets \citep{mayor95, exoplanet95, exoplanets03}, and following the {\it Kepler} mission \citep{KEPLER10}, the focus of exoplanetary research has grown from individual planet detections to studies of the statistical distributions of planets. In addition, the growth and quality of available data now enables us to study and characterize individual exoplanets in more quantitative detail.

In the search for exoplanets, much interest is now directed to planets orbiting their host star in the Habitable Zone (HZ). The commonly used definition of the HZ --- the potential for a planet to have liquid water on its surface as a function of host stellar luminosity and  planetary orbital separation \citep[e.g.,][]{Kasting93} --- limits the physics involved in characterizing exoplanets to the radiation forcing applied at the top of a planetary atmosphere.  Among other effects that can impact habitability are atmospheric loss due to intense EUV/X-ray radiation \citep[focusing on hot jupiters, e.g.,][]{Lammer03,Baraffe04, Yelle04,Baraffe06,Garcia-Munoz07,Murray-clay09,Adams11,Trammell11,OwenJackson12}, cloud coverage and albedo, greenhouse gases, and atmospheric circulation \citep[e.g.,][]{Tian05,Cowan11,Heller11,vanSummeren11,Wordsworth13}. 

Despite the ever more detailed characterization of the atmospheres of exoplanets, very few studies have investigated their space environment and its possible physical impact. Of particular importance is the ability of a close-in planet to sustain an atmosphere at all. As an example from our own solar system, it is believed that the Martian atmosphere has been significantly eroded and evaporated over time \citep[e.g.,][with references therein]{Lammer13}. A number of studies have investigated the interaction between the atmospheres of close-in planets and the surrounding plasma using hydrodynamic particle models, where they used scaled stellar wind parameters \citep{Erkaev05,Kislyakova13,Kislyakova14}, or scaled Coronal Mass Ejection (CME) parameters \citep{Khodachenko07,Lammer07} as the input at the top of their models. All of these studies have found high escape rates of different forms from close-in planets, partially due to the intense EUV radiation.

\citet{Sterenborg11} employed detailed magnetohydrodynamic (MHD) wind models tailored to the case of the young, active Sun to show that 
the magnetopause standoff distance of the Archean Earth could have been as small as 4.25 Earth radii ($R_E$) in the face of a much more vigorous solar wind, compared with the present-day distance of $10.7 R_E$ \citep[see also][]{Tarduno10}.  Similarly compressed planetary magnetospheres might be expected for exoplanets around other stars of high magnetic activity \citep[e.g.,][]{Khodachenko07,Johansson09,Johansson11a,Johansson11b,Vidotto13,See14}. More recently, \cite[][Co14 hereafter]{Cohen14}, studied the interaction of stellar winds with Earth-like, magnetized exoplanets orbiting M-dwarf stars at habitable zone distances. They found that such close-in planets can transition between sub- and super-Alfv\'enic plasma environments along their orbits, and that this dynamic transition leads to a major heating at the top of the planetary atmosphere. 

The interaction of non-magnetized planets with a stellar wind is very different to such an interaction of magnetized planets.  In the former case, the stellar wind directly interacts with the atmosphere and not with the planetary magnetic field and magnetosphere \citep{RussellKivelson95}. In the solar system, such an interaction occurs at Mars, which has a very thin atmosphere and weak crustal field in its southern hemisphere \citep[e.g.,][]{Ma04a}.  Direct interaction between the solar wind and the atmosphere also occurs at Venus. In the Venusian case, the atmosphere is thick enough to balance the pressure from the solar wind \citep{VenusBook83}. 

Modeling the interaction of non-magnetized bodies with a flow of magnetized gas is very complicated, because an accurate approach needs to account for the atmospheric chemistry, ionization and radiation processes, and the atmospheric acceleration processes.  All of these processes define the planetary mass source that resists the stellar wind plasma flow, and the induced magnetosphere created as a result of this interaction.  A number of models have been developed to study this type of problem, including  the interaction of the solar wind with Mars \citep[e.g.,][with references therein]{Ma04a,Ma07a}, and the interaction of Titan with the Kronian (Saturn) magnetosphere \citep[e.g.,][with references therein]{Ma04b,Ma06}. 

\cite{Ma13} (Ma13 hereafter) have studied the interaction between the Venusian atmosphere and the solar wind. They focused on studying the shock location in front of the planet for solar minimum and solar maximum conditions, and on the overall three-dimensional structure of the nearby space environment. Their model compared well with observations. 

In this paper, we use the model by Ma13 to study the interaction between Venus-like exoplanets around M dwarfs and the stellar wind of their host star.  Since the parameters describing exoplanetary systems are currently only loosely constrained, we leave many features of the model unchanged.  Here, we focus on close-in planetary orbits and the competing effects of the enhanced upstream stellar wind and the increased EUV radiation. These are the two main changes we apply to the original modeling work of Ma13. 

The structure of the paper is as follows. In the next section, we describe the numerical model and the overlaying assumptions we make. We describe the results in Section~\ref{sec:Results} and discuss their consequences in Section~\ref{sec:Discussion}. We summarize our conclusions in Section~\ref{sec:Conclusions}.


\section{NUMERICAL SIMULATIONS}
\label{sec:Model}

\subsection{SIMULATIONS SETUP}
\label{sec:GM}

All the simulations presented here were done using the generic {\it BATS-R-US} MHD code \citep{powell99,Toth12}, in a Cartesian geometry. The simulation domain is defined as $-40R_v<X,Y,Z<40R_v$, with Venus radius, $R_v=6052\;km$. The simulation coordinate system is defined as follows. The X axis is defined as the line joining the centers of the planet and star, the Z axis is perpendicular to the ecliptic plain, and the Y axis completes the right-hand system. We use initial mesh refinement to generate a grid structure with high resolution concentrated around the planet, with a grid cell size of 60 km between the surface and an altitude of about 600 km. This high resolution enables us to better resolve the induced ionosphere of the non-magnetized planet. The stellar wind upstream conditions are imposed as fixed boundary conditions at the box side facing the star and the simulation is run until a steady-state is achieved using the local time stepping method \citep[to accelerate convergence time][]{Toth12}. In the case of sub-Alfv\'enic stellar wind conditions, we impose a float boundary condition for the pressure in order to mimic the two-way interaction between the incoming flow and the flow in the simulation domain. 

All the simulations presented here were performed on the Smithsonian Institute's HYDRA cluster.

\subsection{MHD EQUATIONS}
\label{sec:GM}

The model solves the following set of equations.\\

\noindent{\it Multi-species continuity equations:}
      
\begin{eqnarray}
&\frac{\partial \rho_i}{\partial t}+\nabla\cdot(\rho_i \mathbf{u})=S_i-L_i,& \nonumber \\
&S_i=m_iN_i \left( \nu_{ph,i}+\nu_{imp,i} + \sum\limits_{s} k_{si}n_s \right),& \nonumber  \\
&L_i = m_in_i \left( \alpha_{R,i}n_e + \sum\limits_{t} k_{it}n_t \right),&
\label{Continuity}
\end{eqnarray}
where the index $i$ stands for the different ion species used for Venus - H$^+$, O$^+$, O$_2^+$, and CO$_2^+$. 

Here, $\rho_i$ are the mass densities of the different ions and the corresponding number densities are denoted by $n_i$, $m_i$ is the mass of the particular ion, and $N_i$ is the number densities of the neutral sources, CO$_2$ and O. The ion source term, $S_i$, depends on the photoionization rate, $\nu_{ph,i}$, the impact ionization rate, $\nu_{imp,i}$, and the charge exchange rate constant, $k_{si}$ (the index $s$ sums over all the ions). The ion loss term, $L_i$, depends on the charge exchange rate constant, $k_{it}$ (the index $t$ sums over all the neutrals), and the recombination rate constant, $\alpha_{R,i}$. The rates and rate constants of the chemical reactions considered here are summarized in Table~\ref{tab:t1}. 

Once the number densities of the different ions are determined, the total plasma density is defined as $\rho=\sum\limits_i \rho_i$. This plasma density is then used in the rest of the model equations.\\

\noindent{\it Momentum equation:}

\begin{eqnarray}
&\frac{\partial(\rho\mathbf{u})}{\partial t} +\nabla\cdot\left( \rho\mathbf{u}\mathbf{u} +p\mathbf{I} +\frac{B^2}{2\mu_0}\mathbf{I} - \frac{1}{\mu_0}\mathbf{B}\mathbf{B}\right)& \nonumber \\
& = \rho \mathbf{G} -\sum\limits_{i}\rho_i \sum\limits_{t}\nu_{it}\mathbf{u} -\sum\limits_{i}L_i\mathbf{u}.& 
\label{Momentum}
\end{eqnarray}
Here, $\mathbf{u}$ is the (single fluid) plasma velocity vector, $p$ is the pressure, $\mathbf{B}$ is the magnetic field vector, $\mathbf{G}$ is the gravitational acceleration, $\mu_0$ is the magnetic permeability in vacuum, $\mathbf{I}$ is a unit matrix, and $\nu_{t,i}=4\cdot10^{-10}\{ [O]+[CO_2] \}\;s^{-1}$ is the collision frequency between ion $i$ and neutral $t$ \citep{SchunkNagy80}.\\

\noindent{\it Induction equation:}

\begin{eqnarray}
&\frac{\partial \mathbf{B}}{\partial t}+\nabla\cdot(\mathbf{u}\mathbf{B}-\mathbf{B}\mathbf{u})=\nabla\times\left( \frac{1}{\mu_0\sigma_0}\nabla\times B \right).& 
\label{Induction}
\end{eqnarray}

This equation describes the impact of the ambient plasma on the magnetic field, where the right-hand side term describes the effect of the induced magnetosphere, with conductivity $\sigma_0$. $\sigma_0$ depends on the electron mass, $m_e$, the electron number density, $n_e$, the electron charge, $e$, and the atmospheric collision frequencies between electrons and ions ($\nu_{ei}$), and between electrons and neutrals ($\nu_{en}$):

\begin{equation}
\sigma_0=\frac{n_e e^2}{m_e(\nu_{ei}+\nu_{en})}.
\label{SigmaZero}
\end{equation}

The collision frequencies are taken from \cite{SchunkNagy80} and they both depend on the electron temperature, $T_e$. $\nu_{ei}$ also depends on the electron density, $n_e$, while $\nu_{en}$ depends on the neutral densities, [O] and [CO$_2$]. \\

\noindent{\it Energy equation:}

\begin{eqnarray}
&\frac{\partial \varepsilon}{\partial t} + \nabla \cdot \left[ \mathbf{u} \left( \varepsilon + p + \frac{1}{2\mu_0}B^2 \right)  - \frac{1}{\mu_0}(\mathbf{B}\cdot\mathbf{u})\mathbf{B} +B\times\frac{\nabla\times B}{\mu_0\sigma_0}   \right]& \nonumber \\
&= \rho\mathbf{u}\cdot\mathbf{G} -\sum\limits_{i}\sum\limits_{t}\frac{\rho_i\nu_{it}}{m_i+m_t}\left[ 3k(T_n-T_i)-m_iu^2 \right]& \nonumber \\
&-\frac{1}{2}\sum\limits_{i}L_i\mathbf{u}^2 + \frac{k}{\gamma-1}\sum\limits_{i}\left( \frac{S_iT_n-L_iT_i}{m_i} - \frac{\rho_i}{m_i}\alpha_{R,i}n_eT_e\right),&
\label{energy}
\end{eqnarray}
where $\varepsilon= \frac{1}{2}\rho u^2 + \frac{1}{\gamma-1}p + \frac{1}{2\mu_0}B^2$ is the energy density, $\gamma=5/3$ is the ratio of specific heats, $m_t$ is the mass of neutral $t$, $k$ is the Boltzmann constant, and $T_n=1500K$ is the temperature of the newly created ions. In this model we assume that $T_e$ and the ion temperature, $T_i$, are the same and that they are both equal to half of the plasma temperature, $T_p=kp/m_p\rho$.


\subsection{STELLAR WIND INPUT} 
\label{sec:StellarWind}

For the upstream stellar wind conditions of planets orbiting an M-dwarf star, we follow the work presented in Co14. The stellar wind conditions of EV Lac, an M3.5~V  star, were obtained using an MHD model for the stellar corona and wind. This model was driven by available surface magnetic field data of the star \citep[][see Co14 for more details on the coronal model]{Morin08}. Once the stellar wind solution is obtained, the particular conditions are extracted at the orbits of three potentially habitable planets observed by {\it Kepler} - Kepler Object of Interest (KOI) 2626.01, 1422.02, and 854.01 \citep{Dressing13}. These planets are located at distances of 0.06, 0.085, and 0.15 AU, respectively. For each of these planets, we study two cases in which the planets reside in sub- or super-Alfv\'enic stellar wind conditions. For reference, we simulate Venus itself at an orbit of 0.72~AU and a "Venus at the orbit of Mercury" case at 0.39~AU. For these two cases, the stellar wind is super-Alfv\'enic, but we study a case of fast and a case of slow stellar wind. We also assume here that the interplanetary magnetic field (IMF) has a pure $B_z$ component. This is due to the fact that the planet does not have an internal field, and that the IMF direction does not seem to affect the O$^+$ escape from Venus \citep{Masunaga13}. The upstream stellar wind conditions used for all cases are summarized in Tables~\ref{tab:t2} and \ref{tab:t3}.  

\subsection{ENHANCED STELLAR EUV FLUX} 
\label{sec:EUVflux}

In Ma13, the changes in EUV radiation over the solar cycle were considered with two cases for solar minimum and solar maximum. Their model also included the radiation change with Solar Zenith Angle (SZA).  

Planets that reside in close-in orbits are expected to receive enhanced amounts of X-ray and EUV radiation compared to the Earth, Venus, and even Mercury. In order to consider the effect of this enhanced EUV radiation, we scale the photoionization rates of the solar maximum case from Ma13 with the distance square. The scaling parameter of Venus's EUV flux, $F_{\venus}$, for the different cases is shown in the last column of Tables~\ref{tab:t2} and \ref{tab:t3}.

We stress that here we expect the enhanced EUV radiation to simply increase the creation of ions via photoionization processes. In our simplified model for the atmosphere, we do not account for the overall atmospheric heating due to the close proximity to the star (since we are considering the habitable zones of M dwarfs, the optical and IR heating will be similar to the venus-earth-mars case), nor to the possible change in the electron temperature as a result of the creation of super-thermal electrons (i.e., we do not scale $T_e$, $T_i$, and $T_n$ with the EUV flux). These effects can potentially lead to charge separation as the result of the change in ion and electron scale heights, and to additional acceleration and escape of atmospheric ions \citep[i.e., the polar wind,][]{Cohen12}. We leave the implementations of these processes in the model for future work.  

\subsection{ATMOSPHERIC CHEMISTRY, THE NEUTRAL ATMOSPHERE, AND INNER BOUNDARY CONDITIONS} 
\label{sec:StellarWind}

The model assumes a neutral planetary atmosphere source composed of Carbon dioxide, which dominates below 160~km, and Oxygen, which dominates above that height (at Venus). The atmospheric neutral densities assumed here are \citep[the solar maximum case in Ma13,][]{FoxSung01}:
\begin{eqnarray}
&[CO_2]=1.0\cdot 10^{15}\cdot e^{-(z-z0)/5.5}\;cm^{-3}& \nonumber \\
&[O]=2.0\cdot 10^{11}\cdot e^{-(z-z0)/17}\;cm^{-3},&
\label{NeutralDensities}
\end{eqnarray}
where the altitude is in km, and $z_0=100$ km is the inner boundary of the model.

As seen from Eq.~\ref{Continuity}, we assume that the atmospheric chemistry is driven by these two neutral species, and it is governed by the photoionization, impact ionization, charge-exchange, and recombination processes described in Table~\ref{tab:t1}. In future work, we plan to implement other species in order to study exoplanetary atmospheric environments.

The densities for the ion species at the inner boundary are specified by the photochemical equilibrium. The inner boundary is set to absorb the velocity and magnetic field vectors. This guarantees that the magnetic field diffuses through the boundary with the help of the low conductivity (as a result of the strong ion-neutral coupling at these low altitudes). The plasma temperature is assumed to be twice the (Venusian) neutral temperature of 200 K. Here we also neglect a potential increase of the neutral temperature due to close proximity to the star (assuming that such an increase is possible considering the actual stellar radiation of M-dwarfs). 

\clearpage
\section{RESULTS}
\label{sec:Results}

\begin{figure*}[h!]
\centering
\includegraphics[width=3.in]{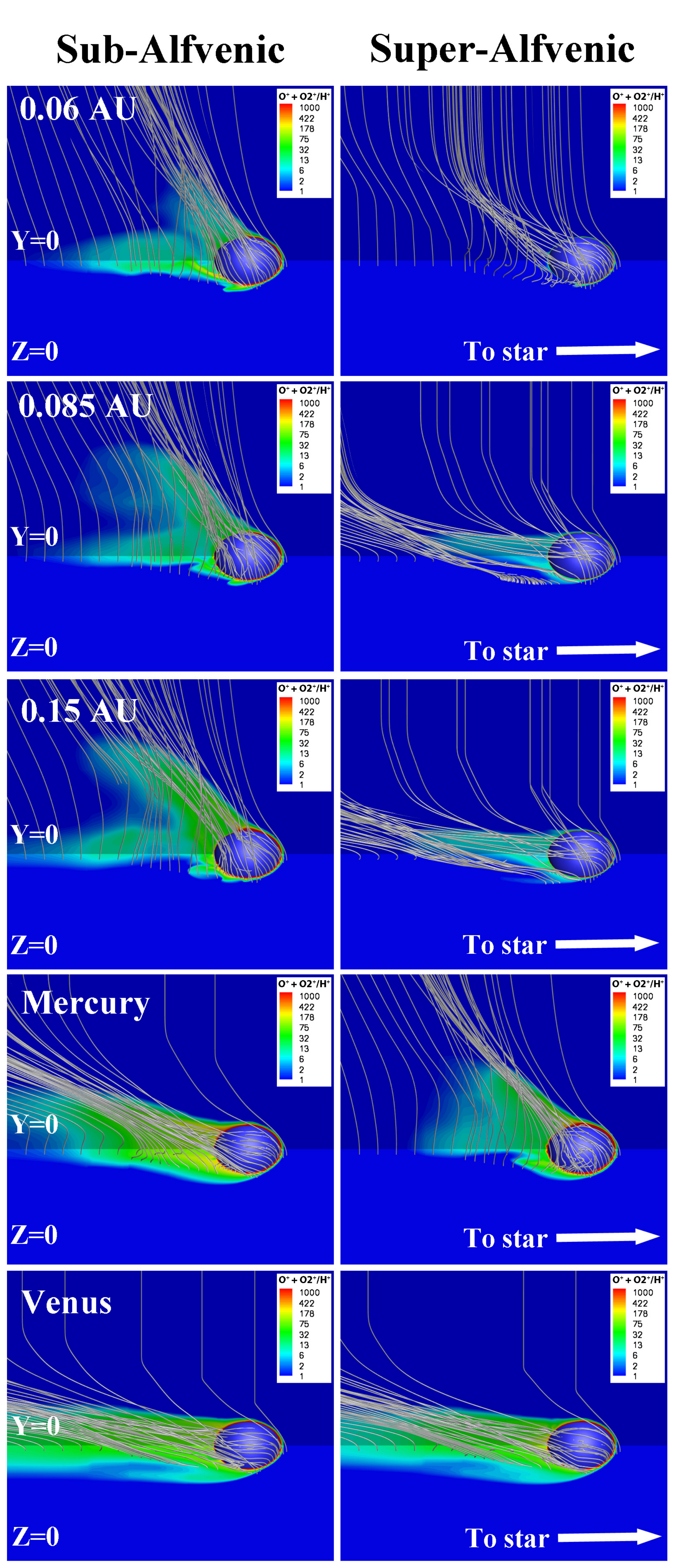}
\caption{The solutions for all the cases shown on the Y=0 and Z=0 plains. The view is from the side where the stellar wind is coming from the right (the direction to the star). The left column shows the sub-Alfv\'enic cases and the right column shows the super-Alfv\'enic cases; the cases change from the closest planet (top) to the farther (Venus, bottom). Color contours are of the ratio of the Oxygen ion density (O$^+$+O$_2^+$) to the H$^+$ density. Selected magnetic field lines are shown as white lines.}
\label{fig:f1}
\end{figure*}

Figure~\ref{fig:f1} shows the model solutions in the Y=0 and Z=0 planes for all cases.  The star is on the right, and the colored contours indicate the ratio of ionized Oxygen (O$^+$+O$_2^+$) to H$^+$ densities. The planet is represented by a sphere and selected magnetic field lines and also shown. 

It can be seen that in all the cases, parts of the planetary environment are dominated by the oxygen ions over the hydrogen ions by factors of a few tenths (and up to a thousand near the inner boundary). The exception is the case for 0.06~AU with a super-Alfv\'enic wind, which remains dominated by the stellar wind hydrogen population. In all other cases, the lower parts of the atmosphere are completely occupied by oxygen, as are large regions in the planetary wake (these are occasionally referred to as the "induced magetosphere" and "induced magnetotail"). 

A notable feature is that the oxygen-dominated regions are largely confined to the equatorial plain in most of the super-Alfv\'enic cases, while these regions extend to lobes above and below the planet in the sub-Alfv\'enic case. The super-Alfv\'enic Mercury case also shows similar structure due to the low Alfve\'nic Mach number (close to 2) in this case. The structure that eappears in the sub-Alfv\'enic cases is typical for a body that moves in ambient plasma with sub-Alfv\'enic speed, where the ambient field is perturbed by the body via a unique current system known as "Alfv\'en Wings" \citep{Neubauer80,Neubauer98}. In this topology, which has been studied in the context of the Jovian moons of Io and Ganymede \citep{Combi98,Linker98,Jacobsen07,Ip02,Kopp02,Jia08}, two standing ``lobes" are created above and below the body. In our simulations, these lobes are filled with the oxygen ions that originate from the planetary atmosphere.

\begin{figure*}[h!]
\centering
\includegraphics[width=6.in]{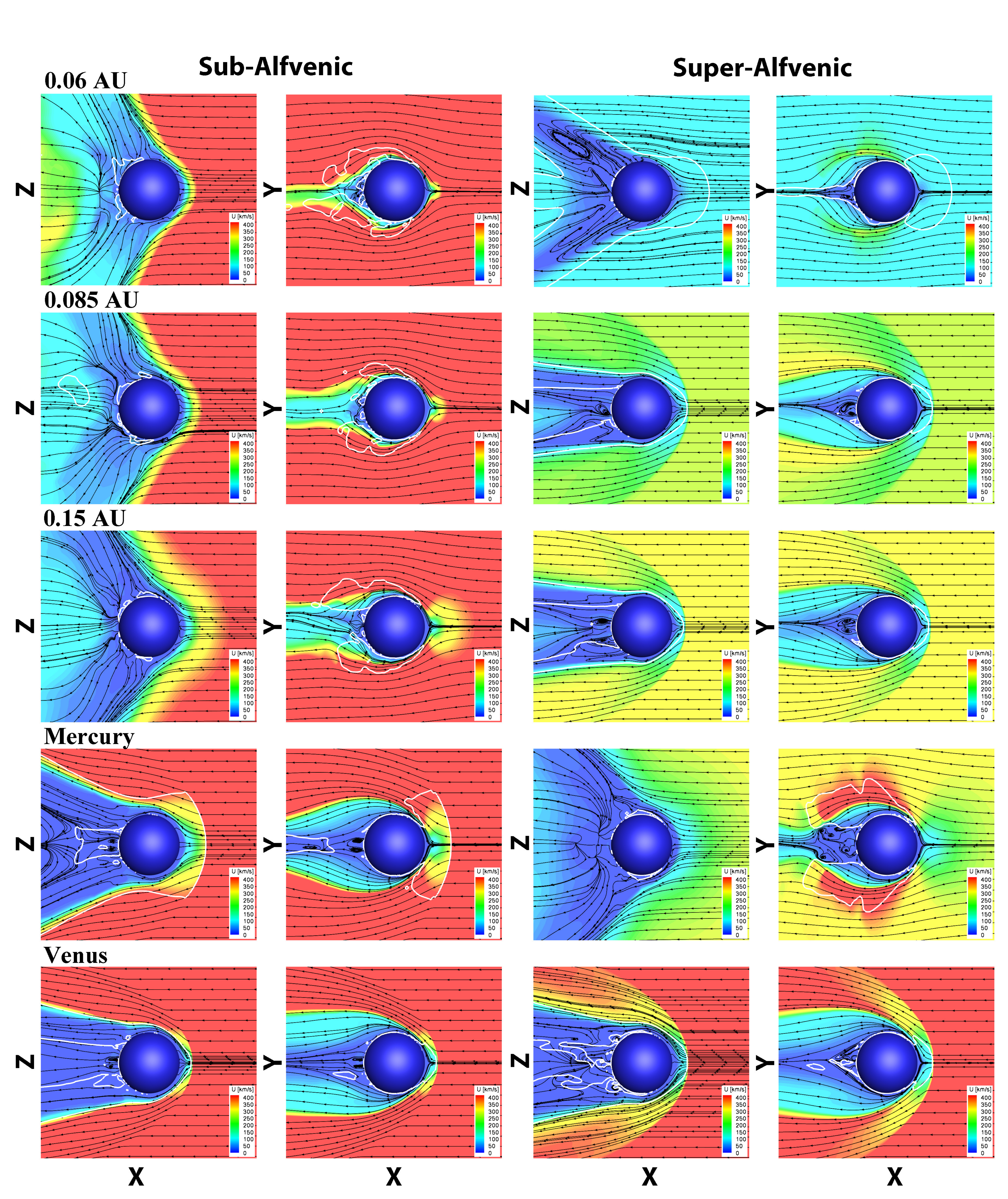}
\caption{The plasma flow pattern for all the cases. Each pair of panel shows the meridional (X-Z) plain and the equatorial (X-Y) plain. The star is on the right. Color contours show the plasma velocity magnitude and the black lines show streamlines. The solid white line represents the contour for $M_a=1$.}
\label{fig:f2}
\end{figure*}

Figure~\ref{fig:f2} shows the velocity distribution for all the cases on the Y=0 (meridional plain) and the Z=0 (equatorial plain), together with the contour line at which the Alfv\'enic Mach number equals one. One can see that there is no bow shock in front of the planet for the sub-Alfv\'enic cases, and that the stellar wind is directly slowed down to speeds of less than 100 km/s at a altitude of about 600-800 km at the sub-stellar point (it is even lower in altitude at the planetary flanks). In the super-Alfv\'enic cases, a shock is formed at altitudes of at least twice this height, but the stellar wind incoming speeds are rather high (about 100 km/s) at much lower altitudes. 

Another representation of the Alfv\'en wing topology is the vertically extended planetary wake with very low speeds seen in Figure~\ref{fig:f2} Y=0 plains for the sub-Alfv\'enic cases and the super-Alfv\'enic cases with low Alfv\'enic Mach number (0.06AU and Mercury). The super-Alfv\'enic cases with strong shock (i.e., high Alfv\'enic Mach number) show a wake that is much more confined towards the equator. Overall, all cases show regions around the planet with a very low flow speed, which indicates a stellar wind free zone.  

The velocity pattern in the sub-Alfv\'enic cases shows that the flow is deflected to very large angles as the result of the standing vertical lobes. In the Super-Alfv\'enic cases, the flow simply goes around the planet, generating a slightly turbulent wake behind the planet in the slow speed tail region.

\begin{figure*}[h!]
\centering
\includegraphics[width=7.in]{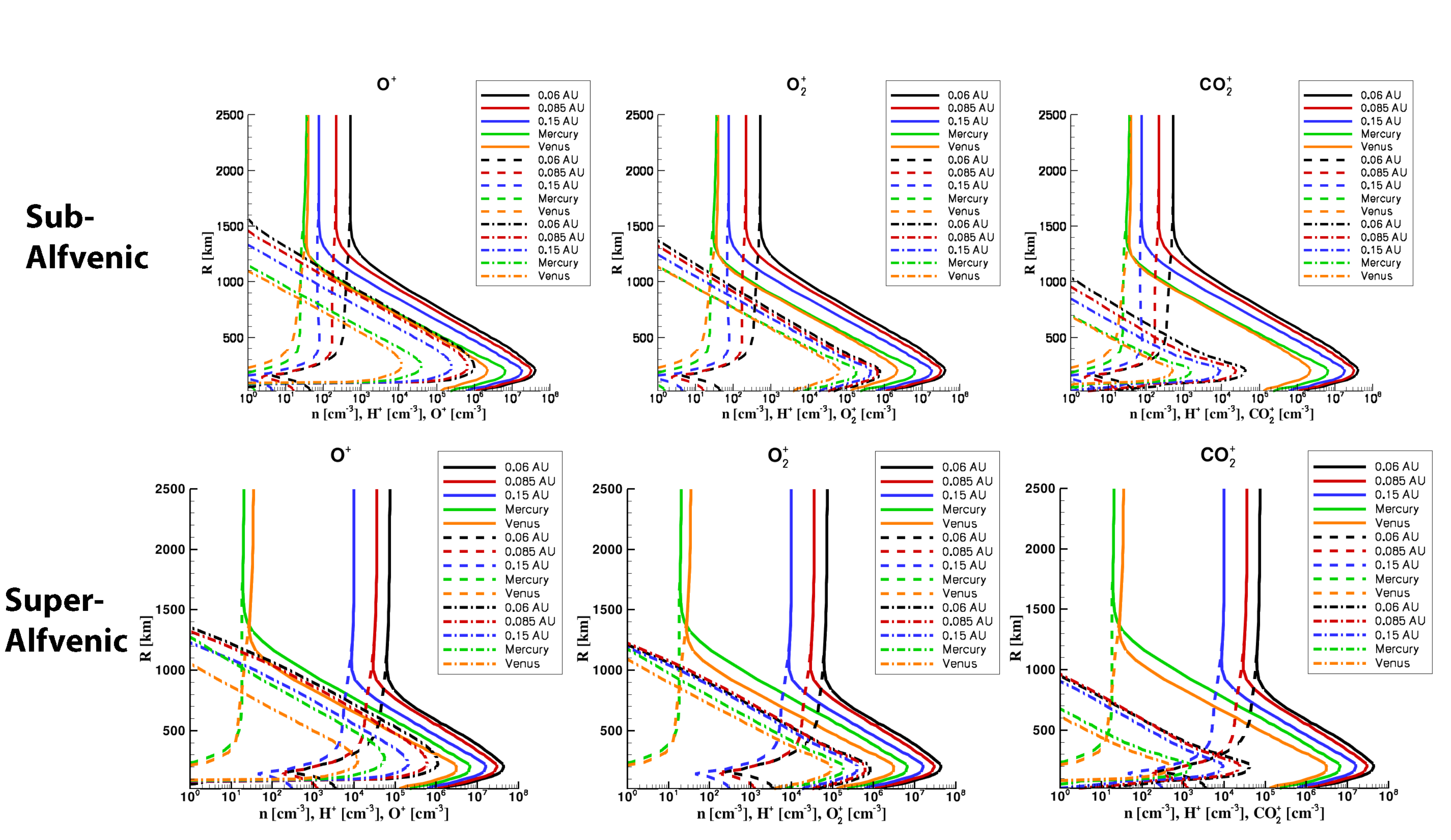}
\caption{Atmospheric density structure for the sub-Alfv\'enic (top) and the super-Alfv\'enic (bottom) cases extracted from the planetary surface along the sub-stellar line (X axis). Each panel shows the total density (solid), H$^+$ density (dashed), and ion density (dotted-dashed) as a function of altitude. The columns are for the ions of O$^+$ (left), O$_2^+$ (middle), and CO$_2^+$ (right).}
\label{fig:f3}
\end{figure*}

Figure~\ref{fig:f3} shows the total ion density, and the densities of the different ion species, as a function of altitude for all the cases. The densities are extracted from the planetary surface along the sub-stellar line (X axis). The total ion density is dominated by H$^+$ down to an altitude of 1000--1500 km, while below this height the atmospheric ions begin to dominate the ion density and peak around 300 km. The enhanced creation of atmospheric ions as a result of the increased EUV flux is clearly reflected in an increase in ion densities as a function of decreasing orbital distance. The plots also show that the altitudes at which the densities of the atmospheric ions are comparable with the density of stellar wind H$^+$ are located lower in the atmosphere for the super-Alfv\'enic conditions compared to the sub-Alfv\'enic conditions (for the three close-in cases). This is due to the fact that the stellar wind density is much higher for the super-Alfv\'enic cases.  


\section{DISCUSSION}
\label{sec:Discussion}

The results of our simulations show that for close-in Venus-like exoplanets, a very strong, direct interaction between the stellar wind and the upper atmosphere occurs at an altitude of no more than about one planetary radius (a few thousand kilometers). This is in contrast to the few planetary radii for the case of magnetized planets presented in Co14. 

\subsection{TOPOLOGY OF THE NEARBY SPACE ENVIRONMENT}
\label{sec:Environment}

The results show a quite distinct topology of the plasma environment near the planet for the sub- and super-Alfv\'enic stellar wind upstream conditions. This suggests that the space environment of close-in Venus-like planet can be very dynamic as these planets pass between sub- to super-Alfv\'enic plasma environments along their orbit (as demonstrated dynamically in Co14 but excluded here). Therefore, through planetary orbital motion alone (regardless of any stellar activity), the stellar wind can affect the planetary atmosphere in terms of different forms of energy deposition, such as Ohmic dissipation, Joule heating, and the generation of gravity waves. All these are expected due to the dynamic change of the induced ionosphere-magnetosphere.

\subsection{STELLAR WIND PENETRATION}
\label{sec:SWpenetration}

\begin{figure*}[h!]
\centering
\includegraphics[width=5.in]{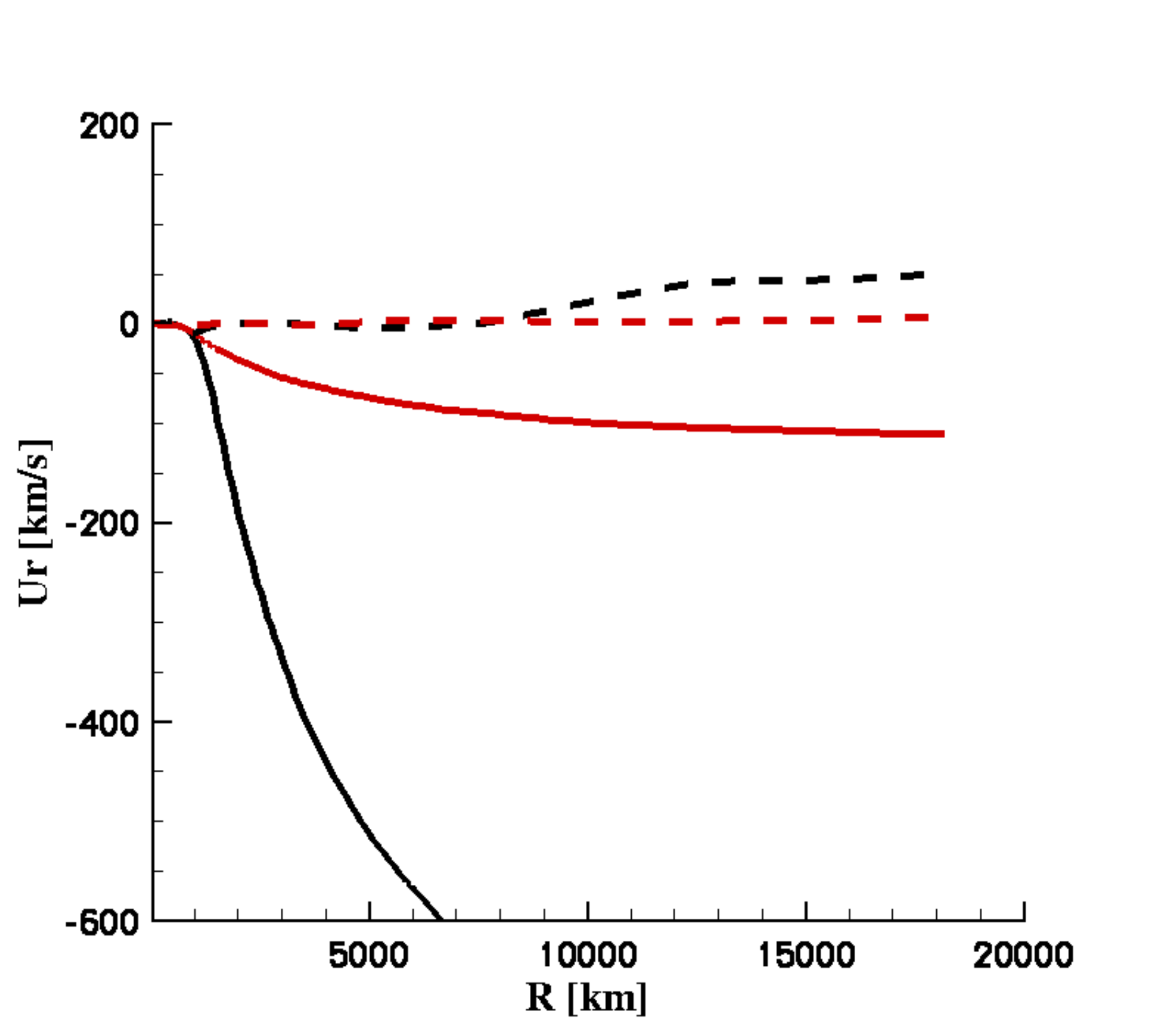}
\caption{Radial velocity as a function of altitude for the nearest non-magnetized cases calculated here (solid) and for the nearest magnetized cases calculated in Co14 (dashed). All cases located at 0.06~AU. Black lines represent sub-Alfv\'enic cases and red lines represent super-Alfv\'enic cases.}
\label{fig:f4}
\end{figure*}

 Figure~\ref{fig:f4} shows the plasma radial speed as a function of altitude up to about 3 planetary radii for the magnetized planets from Co14 and the non-magnetized planets studied here. The plot shows the cases corresponding to an orbital  distance of 0.06~AU. A negative sign of this speed component represents an inflow, while a positive sign represents an outflow. The plot clearly shows that for the magnetized planet cases, the inflow is already very weak far from the planet and there is even weak outflow in the sub-Alfv\'enic case. In the case of non-magnetized planets, the inflow slowly decreases for the super-Alfv\'enic case (this is the post-shock flow), and sharply decreases for the sub-Alfv\'enic case. 
 
\begin{figure*}[h!]
\centering
\includegraphics[width=7.in]{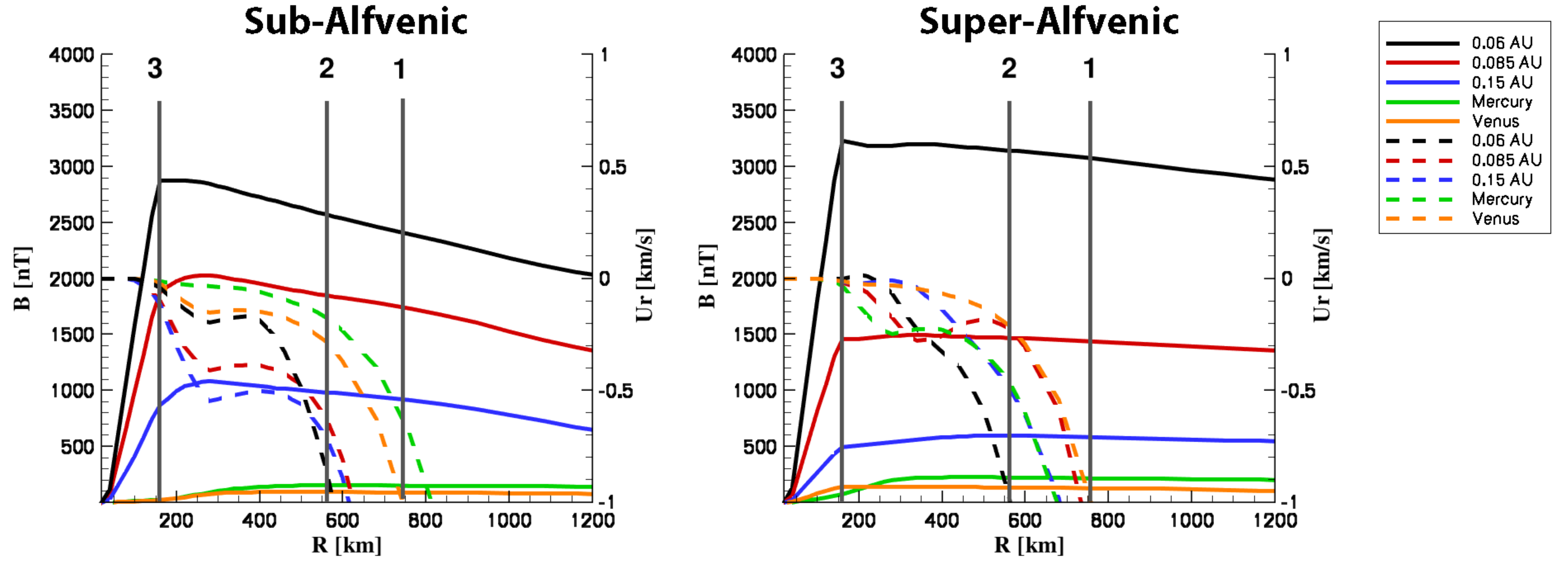}
\caption{Magnetic field magnitude (solid lines) and radial velocity (dashed lines) as a function of altitude for all cases are extracted along the sub-stellar line (X axis).}
\label{fig:f5}
\end{figure*}

Figure~\ref{fig:f5} shows the radial speed and the magnitude of the magnetic field as a function of altitude for all the non-magnetized Venus-like cases. The values are extracted from the planetary surface along the sub-stellar line (similar to Figure~\ref{fig:f3}). The scale for the radial speed is designed to demonstrate the point at which the incoming wind gets to a complete stop (inflow speed of less than 1 km/s). In both the sub- and super-Alfv\'enic cases, line 1 marks the point where the wind stops in the cases of Venus, and line 2 marks the point where the wind stops in the case of the planet located at 0.06 AU. In both cases, the altitude difference is about 200 km and the wind penetrates up to an altitude of less than 600 km above the planetary surface.  

The magnetic field slowly increases as the result of the plasma compression in front of the planet, until a maximum is attained, and then starts to decrease to zero at an altitude of around 180~km for the case of distance of 0.06~AU (marked by line 3). This point, which marks the decline of the magnetic field and represents the magnetic field-free planetary atmosphere is the same for all the cases of super-Alfv\'enic wind conditions (with the exception of the Mercury case which is almost sub-Alfv\'enic). In the sub-Alfv\'enic cases, this point moves outward as the planet gets further from the star, up to about 350~km for the Mercury and Venus cases. This difference is probably due to the fact that the stellar wind in the sub-Alfv\'enic cases is much less dense than the super-Alfv\'enic one. Therefore, the compression region of the incoming stellar wind field, as well as the ability of the field to diffuse towards the atmosphere (as the result of increased ion/electron density and their effect on the conductance and collision frequencies), are more sensitive to the Oxygen outflow in the sub-Alfv\'enic case than the super-Alfvenic case. 

\begin{figure*}[h!]
\centering
\includegraphics[width=7.in]{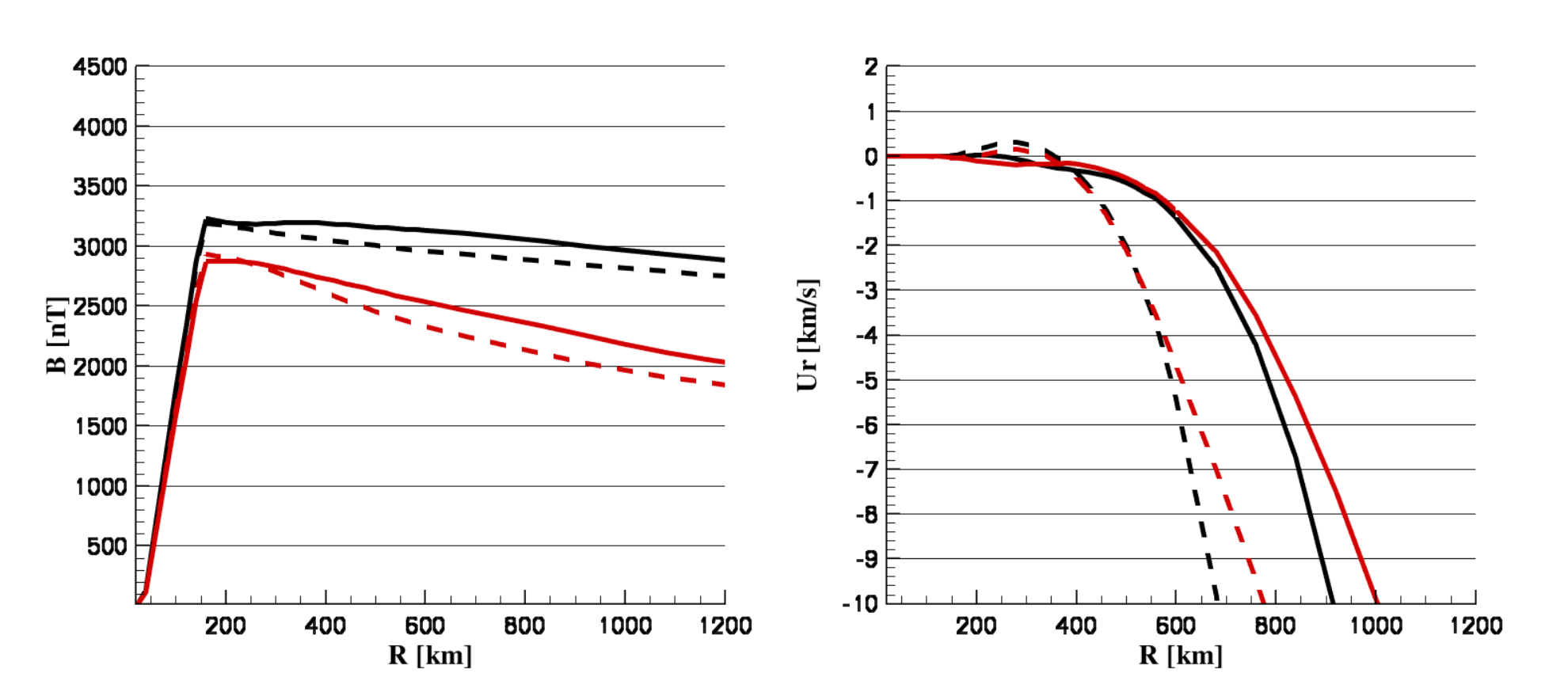}
\caption{Magnetic field magnitude (left) and radial velocity (right) as a function of altitude for the nearest cases at 0.06~AU are extracted along the sub-stellar line (X axis). Black lines represent sub-Alfv\'enic cases and red lines represent super-Alfv\'enic cases. Solid lines represent the cases with EUV flux scaled with distance while dashed lines represent cases with unscaled EUV flux (Venus EUV flux).}
\label{fig:f6}
\end{figure*}

To demonstrate the strong effect of the increased EUV radiation, Figure~\ref{fig:f6} shows the difference in the radial speed as a function of altitude for the case located at 0.06~AU with and without scaling of the EUV radiation with distance. The plot clearly shows that the enhanced radiation increases the production of atmospheric ion flux, which stops the stellar wind about 200 km higher than the case without increase EUV flux. This increased ion flux also results in an enhanced compression of the magnetic field, even though it does not seem to affect the point at which the magnetic field starts to decrease.

\subsection{ATMOSPHERIC ION ESCAPE}
\label{sec:IonEscape}

The model used here separates the densities of the different ions in the continuity equation but it does not distinguish between the different ion velocities and uses a single-fluid plasma velocity in the momentum, induction, and energy equation. Since we do not have information about the particular ion velocities, it is hard to estimate the escape rate of the different ions, in particular the O$^+$ escape rate. In addition and as mentioned in Section~\ref{sec:EUVflux}, the particular model used here does not account for any acceleration of the atmospheric ions via hydrodynamic acceleration \citep[e.g.,][]{Garcia-Munoz07,Tian05}, and the ambipolar electric field \citep{Axford68, BanksHolzer68, Cohen12}, which is the main mechanism to explain the large amount of O$^+$ observed in the Earth's magnetosphere. \cite{Tian09a,Tian09b} even found that the intense EUV radiation increases the amount of ionized CO$_2$ and decrease the amount of neutral CO$_2$. As a result, the atmospheric Infrared (IR) cooling is reduced as well and the atmosphere becomes over-inflated. However, this is a hydrodynamic, one-dimensional model that does not account for day-night energy transfer, electromagnetic effects, and it does not include the dynamic pressure of the incoming stellar wind at the top boundary.

\begin{figure*}[h!]
\centering
\includegraphics[width=7.in]{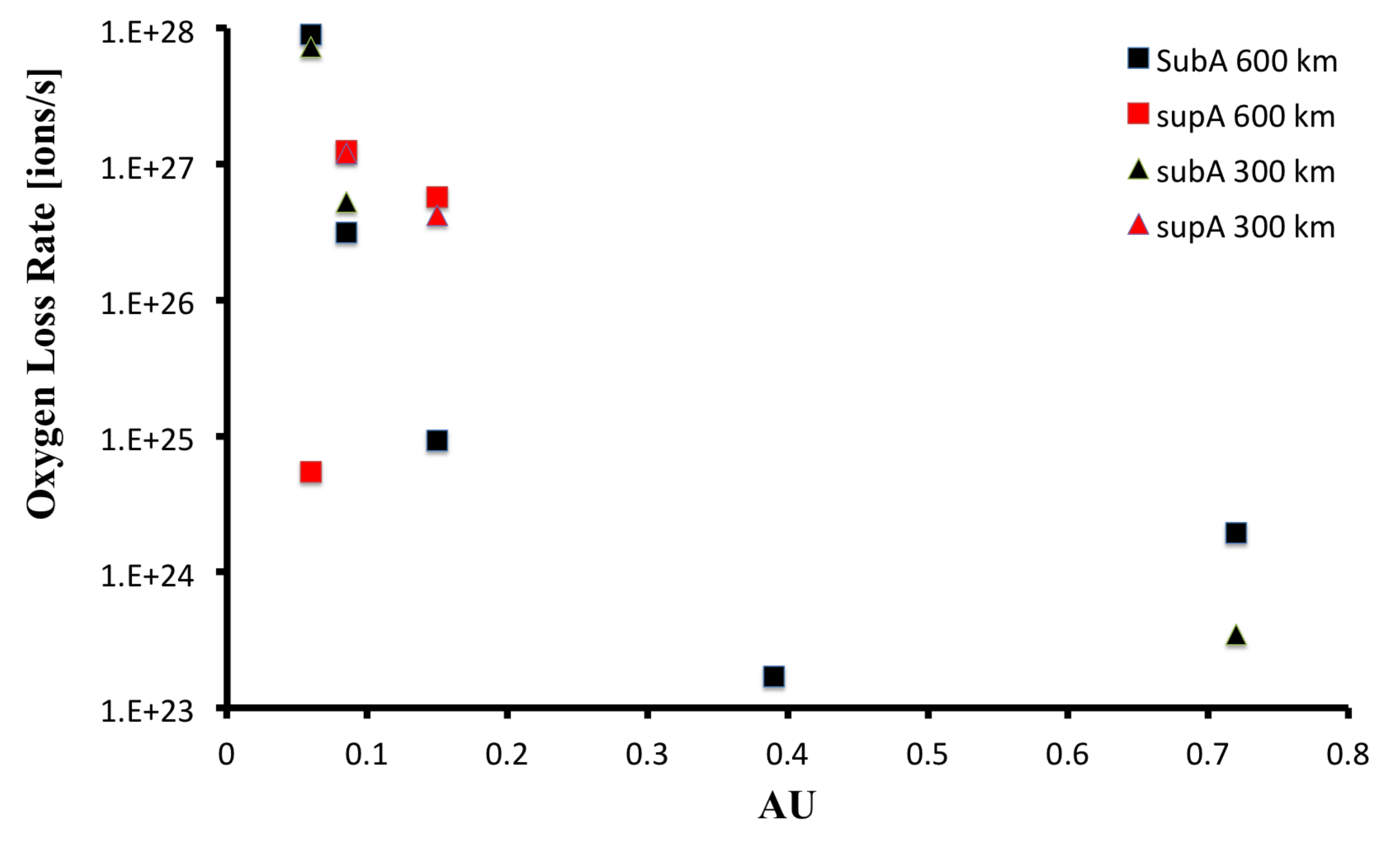}
\caption{O$^+$ escape rate for the different cases as a function of distance from the star. The integration at $300\;km$ did not yield any flux for the 0.06~AU, 0.15~AU, and Mercury cases.}
\label{fig:f7}
\end{figure*}

Nevertheless, we can attempt to estimate a lower limit for the O$^+$ escape rate by calculating the the integral of the mass flux through a sphere at a certain height, and include only the density points at which the radial velocity has both positive sign (i.e., outflow), and magnitude larger than Venus's escape velocity of 10.36 km/s. Figure~\ref{fig:f7} shows the loss rates of O$^+$ for the different cases, integrated on spheres of $r=300\;km$, which is the altitude of the ion density peak, and $r=600\;km$, which is twice that height. The values are not always the same for the two heights as our assumption here does not conserves mass. In addition, some of the integrations result in zero as they do not include any outflow points with speeds above the escape velocity.  

The loss rates for the Venus case are of the order of $10^{24}\;s^{-1}$, and they are comparable with O$^+$ loss rates observed by Pioneer Venus Orbiter and Venus Express \citep[e.g.,][]{Lammer06,Luhmann08,Dubinin11,Fedorov11,Lundin11}. The escape rate is enhanced by 3-4 orders of magnitude for the close-in planets as the result of the enhanced EUV photoionization. The mass loss rates translate to $10^{-15}-10^{-11}\;M_{va}\;yr^{-1}$, with $M_{va}=5\cdot10^{23}\;g$ being the mass of the Venusian atmosphere \citep{BasilevskyHead03}. Interestingly, the O$^+$ loss rates of the close-in planets are similar to those obtained for Venus 4 Gyr ago by \cite{Kulikov06}, who calculated the Venusian loss rate of O$^+$ for different ages of the solar system (taking into account the changes in the stellar radiation environment). The loss rates obtained by our simulations are much lower than the loss rates calculated for atmospheric hydrodynamic escape from hot jupiters \citep[$10^{7}-10^{14}\;g\;s^{-1}$ e.g.,][]{Lammer03,Erkaev05,Penz08b,Murray-clay09,OwenJackson12}. The reason is that hot jupiters lose hydrogen atoms via hydrodynamic blow off. In the case of stellar wind ion pick up \citep[e.g.,][for the Kepler-11 planets]{Kislyakova13,Kislyakova14}, hydrogen-dominated atmospheres are much more expanded compared to CO$_2$ atmospheres, so that the pressure balance between the stellar wind plasma and the hydrodynamically expanding upper atmosphere is much further up compared to the cases presented here.

In the case of a planet located at 0.06~AU, the loss rate is much higher for the sub-Alfv\'enic case comparing to the super-Alfv\'enic case. This is because the strong, dense stellar wind in the super-Alfv\'enic case suppresses the atmospheric Oxygen outflow generated by the enhanced EUV radiation (see top-right panel of Figure~\ref{fig:f1}). The outflow is much more significant for the sub-Alfv\'enic case, at which the stellar wind is less dense. For the cases of planets located at 0.085~AU and 0.15~AU, the loss rate is higher for the super-Alfv\'enic cases comparing to the sub-Alfv\'enic cases due to the slowdown of the stellar wind at the shock in front of the planet. This leaves more room for the atmospheric ions to outflow against the incoming wind.   

At face value, the loss rates estimated by our simulations mean that only a very small fraction of a Venusian-like atmosphere will be lost over the course of a billion years. Therefore, our simulations suggest that there is a good chance that the atmospheres of Venus-like exoplanets orbiting M-dwarf stars are sustainable. Nevertheless, the actual mass-loss rate may be significantly higher than these lower limits. The increase in ion scale-height due to increased EUV photoionization may boost the ion density by an order of magnitude at high altitude. In addition, acceleration of ions at the upper atmosphere as a result of the polar wind can increase the ion velocity by another factor of 10 \citep{Glocer09,Cohen12}. This would mean instead that the planet may lose most of its atmosphere over the course of a billion years. A similar conclusion was drawn by \cite{Lammer07}, who found that close-in terrestrial exoplanets may loose hundreds or even a thousand bars of their atmosphere due to the intense EUV radiation and CME atmospheric erosion, even in the case where the planet has a strong internal magnetic field.

More rigorous estimates of the atmospheric loss would require coupling of the model for the planetary space environment presented here with a model accounting for the structure of the upper atmosphere, that includes the acceleration of the atmospheric ions, the impact of the increased EUV on the atmospheric temperatures and the atmospheric conductance, and a self-consistent photoionization model. Such implementations are needed in order to provide a more accurate assessment of the ability of close-in exoplanets to sustain their atmospheres.  


\section{SUMMARY AND CONCLUSIONS}
\label{sec:Conclusions}

We study the interaction between the atmospheres of Venus-like, non-magnetized exoplanets orbiting M-dwarf stars and the stellar wind using a multi-species MHD model. We focus our investigation on the effect of enhanced stellar wind and enhanced EUV flux as the planetary distance from the star decreases from Venus's orbit at 0.72~AU to close-in orbits located at 0.15, 0.085, and 0.06~AU. 

For the close-in orbits, we study the interaction between sub- and super-Alfv\'enic stellar wind conditions. We find that the two cases result in a very different configuration of the immediate planetary space environment, with the sub-Alfv\'enic case resulting in vertically extended wake, while the super-Alfv\'enic case produces a wake which is confined to the equatorial region. We expect that such dynamic change along the orbit of few days would deposit energy into the planetary atmosphere in the form of heating and the generation of gravity waves.

The stellar wind penetration for the non-magnetized planets studied here is much deeper than the magnetized planets studied in Co14. It reaches altitudes of several hundreds of kilometers, in contrast to thousands in magnetized planets. As the planets reside closer to the star, the increase EUV photoionization leads to an enhanced creation of ions that push the stellar wind stagnation point up by about 200 km compared to a low photoionization rate case. It also increases the compression of the stellar wind magnetic field in front of the planet. 

We estimate a lower limit for the O$^+$ escape rate and find that the planetary atmosphere may be sustainable over the lifetime of the planet. However, the atmospheric escape rate could be much higher when additional acceleration mechanisms for the atmospheric ions are considered. We plan to implement these mechanisms in future work in order to better estimate the ability of Venus-like exoplanets to sustain their atmospheres. 


\acknowledgments

We thank an unknown referee for his/her useful comments. The work presented here was funded by the Smithsonian Institution Consortium for Unlocking the Mysteries of the Universe grant 'Lessons from Mars: Are Habitable Atmospheres on Planets around M Dwarfs Viable?', the Smithsonian Institute Competitive Grants Program for Science (CGPS) grant 'Can Exoplanets Around Red Dwarfs Maintain Habitable Atmospheres?', and by NASA Astrobiology Institute grant NNX15AE05G. Simulation results were obtained using the Space Weather Modeling Framework, developed by the Center for Space Environment Modeling, at the University of Michigan with funding support from NASA ESS, NASA ESTO-CT, NSF KDI, and DoD MURI. The simulations were performed on the Smithsonian Institute HYDRA cluster.  JJD was supported by NASA contract NAS8--03060 to the {\em Chandra X-ray Center} during the course of this research and thanks the Director, B.~Wilkes, for continuing support and encouragement. 



\newpage

\begin{deluxetable}{ccc}
\tabletypesize{\scriptsize}
\tablecolumns{8}
\tablewidth{0pt}
\tablenum{1}
\label{tab:t1}
\tablecaption{Chemical Reactions Used in the Model\,\tablenotemark{a}}
\tablehead{
\colhead{Reaction}  & \colhead{Rate Ceofficient} &  \colhead{Reference} }
\startdata
CO$_2$ + h$\nu$ $\rightarrow$ CO$_2^+$ + e & $3.24\times10^{-6}\;s^{-1}$ & \cite{SchunkNagy04} \\
O + h$\nu$ $\rightarrow$ O$^+$ + e & $1.21\times10^{-6}\;s^{-1}$ & \cite{SchunkNagy04} \\
CO$_2^+$ + O $\rightarrow$ O$_2^+$ + CO & $1.64\times10^{-10}\;cm^{-3}\;s^{-1}$ & \cite{SchunkNagy04} \\
CO$_2^+$ + O $\rightarrow$ O$^+$ + CO$_2$ & $9.6\times10^{-11}\;cm^{-3}\;s^{-1}$ & \cite{SchunkNagy04} \\
O$^+$ + CO$_2$ $\rightarrow$ O$_2^+$ + CO & $1.1\times10^{-9}\;cm^{-3}\;s^{-1}$ for $T_i\le800\;K$;& \cite{FoxSung01} \\
& $1.1\times10^{-9}(800/T_i)^{0.39}\;cm^{-3}\;s^{-1}$ for $T_i>800\;K$& \\
H$^+$ + O $\rightarrow$ O$^+$ +H & $3.75\times10^{-10}\;cm^{-3}\;s^{-1}$ & \cite{SchunkNagy04} \\
O$_2^+$ + e $\rightarrow$ O + O & $1.95\times10^{-7}(300/T_e)^{0.7}\;cm^{-3}\;s^{-1}$ for $T_e\le1200\;K$;& \cite{SchunkNagy04} \\
& $7.38\times10^{-8}(1200/T_e)^{0.56}\;cm^{-3}\;s^{-1}$ for $T_e>1200\;K$ & \\
CO$_2^+$ + e $\rightarrow$ CO + O & $3.5\times10^{-7}(300/T_e)^{0.5}\;cm^{-3}\;s^{-1}$ & \cite{FoxSung01}
\enddata
\tablenotetext{a}{Electron impact ionization is neglected in the calculation, H$^+$ density is from the stellar wind, The neutral hydrogen is neglected.}
\end{deluxetable}

\begin{deluxetable}{ccccccccccc}
\tabletypesize{\scriptsize}
\tablecolumns{8}
\tablewidth{0pt}
\tablenum{2}
\label{tab:t2}
\tablecaption{Sub-Alfv\'enic Stellar Wind Parameters}
\tablehead{
\colhead{$r\;[AU]$ }  & \colhead{$n\;[cm^{-3}]$} &  \colhead{$T\;[K]$} & \colhead{$u_x\;[km/s]$} &  \colhead{$u_y\;[km/s]$} & \colhead{$u_z\;[km/s]$} &\colhead{$B_x\;[nT]$} &\colhead{$B_y\;[nT]$} &\colhead{$B_z\;[nT]$} &\colhead{$M_a$}&\colhead{$F_{EUV}\;[F_{\venus}]$}}
\startdata
$0.06$ (KOI 2626.01)   & $450$ & $1,000,000$  & $-600$ & $0$ & $0$ & $0$ & $0$ & $2,000$ & $0.29$ &$144$\\
$0.085$ (KOI 1414.02) & $160$ & $750,000$     & $-700$ & $0$ & $0$ & $0$ & $0$ & $800$   &$0.5$ &$72$\\
$0.15$ (KOI 854.01)   & $45$    & $500,000$     & $-700$ & $0$ & $0$ & $0$ & $0$ & $250$   &$0.86$ &$23$\\
\,\tablenotemark{*}$0.39$ (Mercury)   & $15$    & $300,000$     & $-700$ & $0$ & $0$ & $0$ & $0$ & $50$   & $2.5$ & $3.4$\\
\,\tablenotemark{*}$0.72$ (Venus)   & $10$    & $250,000$       & $-700$ & $0$ & $0$ & $0$ & $0$ & $10$   & $10.1$  & $1$
\enddata
\tablenotetext{*}{The cases labeled "Venus" and "Mercury" are super-Alfv\'enic.}
\end{deluxetable}

\begin{deluxetable}{ccccccccccc}
\tabletypesize{\scriptsize}
\tablecolumns{8}
\tablewidth{0pt}
\tablenum{3}
\label{tab:t3}
\tablecaption{Super-Alfv\'enic Stellar Wind Parameters}
\tablehead{
\colhead{$r\;[AU]$ }  & \colhead{$n\;[cm^{-3}]$} &  \colhead{$T\;[K]$} & \colhead{$u_x\;[km/s]$} &  \colhead{$u_y\;[km/s]$} & \colhead{$u_z\;[km/s]$} &\colhead{$B_x\;[nT]$} &\colhead{$B_y\;[nT]$} &\colhead{$B_z\;[nT]$} &\colhead{$M_a$}&\colhead{$F_{EUV}\;[F_{\venus} ]$}}
\startdata
$0.06$ (KOI 2626.01) & $35,000$ & $800,000$     & $-200$ & $0$ & $0$ & $0$ & $0$ & $500$ & $3.4$& $144$\\
$0.085$ (KOI 1414.02) & $12,500$ & $450,000$     & $-250$ & $0$ & $0$ & $0$ & $0$ & $250$  &$5.1$ & $72$\\
$0.15$ (KOI 854.01)   & $3,000$    & $200,000$     & $-300$ & $0$ & $0$ & $0$ & $0$ & $100$  & $7.5$& $23$\\
$0.39$ (Mercury)   & $50$          & $300,000$     & $-450$ & $0$ & $0$ & $0$ & $0$ & $50$    &$2.9$ & $3.4$\\
$0.72$ (Venus)  & $10$          & $250,000$       & $-450$ & $0$ & $0$ & $0$ & $0$ & $10$     &$6.5$ & $1$
\enddata
\end{deluxetable}


\end{document}